\documentclass[aps,twocolumn,prb,superscriptaddress,english,longbibliography]{revtex4-2}

\usepackage{amsmath}
\usepackage{amssymb}
\usepackage{amsthm}
\usepackage{makecell}
\usepackage{amsbsy}
\usepackage{amstext}
\usepackage{graphicx}
\usepackage{color}
\usepackage{float}

\usepackage{wasysym}

\usepackage{mathtools}
\usepackage{rotating}
\usepackage{mathtools}
\usepackage{tikz}
\makeatletter
\usepackage{amsfonts}
\usepackage{dcolumn}
\usepackage{bbold,bm}

\def\degree{${}^{\circ}$}

\makeatother
\begin{document}

\title{Exceptional topology in non-Hermitian twisted bilayer graphene}
\author{Yingyi Huang}
\email{yyhuang@gdut.edu.cn}
\affiliation{School of Physics and Optoelectronic Engineering,
Guangdong University of Technology, Guangzhou 510006, China}
\affiliation{Guangdong Provincial Key Laboratory of Sensing Physics and System Integration Applications, Guangdong University of Technology, Guangzhou 510006, China}

\date{\today}

\begin{abstract}
Twisted bilayer graphene (TBG) has extraordinary electronic properties at the magic angle along with an isolated flat band at the magic angle. However, the non-Hermitian phenomena in twisted bilayer graphene remain unexplored. In this work, we study a non-Hermitian TBG formed by one-layer graphene twisted relative to another layer with gain and loss. 
Using a non-Hermitian generalization of the Bistritzer-MacDonald model, we find Dirac cones centered at only the $K_M$ ($K'_M$) corner of the moir\'e Brillouin zone at the $K'$ ($K$) valley deform into rings of exceptional points in the presence of non-Hermiticity, which is different from single-layer graphene with gain and loss, where exceptional rings appear in both $K$ and $K'$ corners of the Brillouin zone. We show that the exceptional rings are protected by non-Hermitian chiral symmetry.
More interestingly, at an ``exceptional magic angle" larger than the Hermitian magic angle, the exceptional rings coincide and form non-Hermitian flat bands with zero energy and a finite lifetime. These non-Hermitian flat bands in the moir\'e system, which are isolated from dispersive bands, are distinguished from those in non-Hermitian frustrated lattices. In addition, we find that the non-Hermitian flat band has topological charge conserved in the moir\'e Brillouin zone, which is allowed for analogs of non-Hermitian fractional quantum Hall states. 

\end{abstract}
\maketitle
\begin{figure}[t]
	\centering	
	\includegraphics[width=0.99\linewidth]{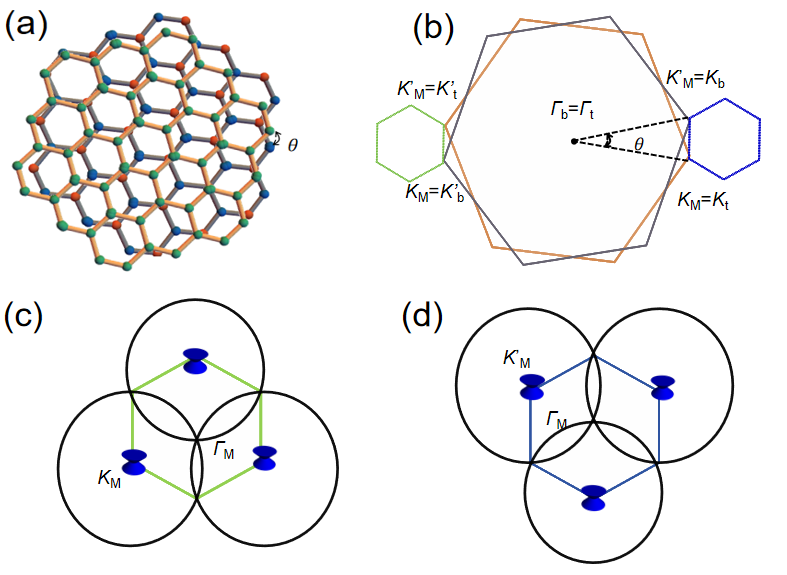}
	\caption{(color online)Schematics of twisted bilayer graphene with onsite gain and loss. (a) Real-space geometrical picture. The gray (orange) honeycomb lattice marks the top (bottom) layer. The red (blue) dots represent the amplifying (lossy) sites in the bottom layer. (b) Momentum-space geometrical picture. The gray (brown) lattice marks the Brillouin zone of the top (bottom) layer graphene. Their expansion around the $K$ ($K'$) valley forms the moir\'e Brillouin zone in green (blue). (c) Moir\'e Brillouin zone physics at the $K'$ valley. Under the effect of on-site gain and loss, Dirac points at the $K_M$ corners (related to $K_b$ points of bottom layer graphene) morph into exceptional 
		rings. At an ``exceptional magic angle", the exceptional rings touch at the $\Gamma_K$ point. (d) Moir\'e Brillouin zone physics at the $K$ valley. Exceptional rings from Dirac points at $K'_M$ are shown. 
	}
	\label{fig:fig1}
\end{figure}
\section{introduction}

Graphene has extraordinary electronic and optical properties in two-dimensional systems ~\cite{neto2009electronic,gonccalves2016introduction}.
These properties relate to its lattice structure, which is directly interconnected with its topological property. In single-layer graphene, the presence of unusual linear band dispersion termed the ``Dirac cone" at $K$ and $K'$ points of the Brillouin zone is derived from its honeycomb lattice geometry. The two Dirac points are related by time-reversal symmetry and thus have opposite Berry phases, which has been confirmed in experiments~\cite{novoselov2005two,zhang2005experimental}. 

When two graphene layers are shifted with each other in rational competing periodicities upon shearing or twisting, they form a moir\'e pattern~\cite{charlier1992tight,mccann2006asymmetry}. One prominent example is a graphene bilayer with a relatively small-angle rotation between the layers~\cite{dos2007graphene}.
 The recent studies on twisted bilayer graphene (TBG) have focused on the extraordinary electronic properties at the magic angle along with an isolated flat band~\cite{Bistritzer_2011,bistritzer2011moire,dos2012continuum,Shallcross_2010,luican2011single}.
 It is related to a variety of exotic phases including the quantum anomalous Hall effect \cite{serlin2020intrinsic,sharpe2019emergent,nuckolls2020strongly,wu2021chern,choi2021correlation}, fractional Chern insulators~\cite{xie2021fractional,  park2023observation, lu2023fractional}, ferromagnetic states~\cite{sharpe2019emergent,saito2021hofstadter,lin2022spin}, nematicity~\cite{cao2021nematicity}, and unconventional superconductivity~\cite{cao2018correlated,cao2018unconventional,yankowitz2019tuning,khalaf2021charged, chatterjee2022skyrmion, torma2022superconductivity}. On the other hand, twisted bilayer graphene has extraordinary topological properties. Since the two Dirac points in moir\'e Brillouin zone (mBZ) in one valley emanate from different layers, their Berry phases can be identical due to the symmetry interlayer hopping term, which is generally not allowed in two-dimensional periodic systems following the Nielsen-Ninomiya theorem~\cite{PhysRevX.9.021013,nielsen1981no}. This is in contrast to single-layer graphene, in which the Berry phases of Dirac cones in opposite valleys have different topological charges.

The topological properties of two-dimensional systems can be generalized to non-Hermitian systems by introducing on-site gain and loss~\cite{ding2022non,ashida2020non,bergholtz2021exceptional}. The presence of non-Hermiticity can transform Hermitian degenerate points, such
as a Dirac-like point or a Weyl point, into a ring of exceptional points~\cite{xu2017weyl,cerjan2018effects}. At an exceptional point, not only do the real and imaginary parts of the eigenvalues degenerate, but also the eigenvectors coalesce.
Many phenomena are unique to non-Hermitian systems, including the non-Hermitian skin effect~\cite{yao2018edge} and non-Hermitian topological classifications~\cite{kawabata2019symmetry,gong2018topological}. Although topological robustness has been shown in one-dimensional moir\'e lattice under stain, the non-Hermitian effect of the twisted moir\'e system with magic angles has been overlooked~\cite{shao2024non,zhang2019high}. 

One important question is whether the existence of flat bands is compatible with non-Hermiticity. Non-Hermitian flat bands have been investigated in non-Hermitian Lieb lattices, kagome lattices, and other non-Hermitian geometrical frustrated lattices~\cite{ge2015parity,chern2015pt,leykam2017flat,qi2018defect,zhang2019flat}, which are protected by either $\mathcal{P}\mathcal{T}$ symmetry or non-Hermitian chiral symmetry~\cite{ge2015parity,leykam2017flat,ge2017symmetry}. However, these non-Hermitian flat bands in simple geometrical frustrated lattices are usually embedded in dispersive bands. From the fact that frustration between Hermitian and non-Hermitian kinetic energy in simple lattices can induce non-Hermitian flat bands, how the non-Hermitian dispersive energy scale interplays with the interlayer tunneling energy scale modulated by the moir\'e pattern becomes interesting.

Another important question is whether the exceptional geometry in mBZ can be topological.
Usually, non-Hermiticity is detrimental to the topology in single-layer graphene.  When the exceptional contours with opposite topological charges merge, the topological charge can be dissipated~\cite{cerjan2018effects}. In single-layer graphene, the annihilation of exceptional points from different valleys has been experimentally observed in a two-dimensional photonic system~\cite{krol2022annihilation}. Since the mBZ valley of TBG has topological properties different from those of single-layer graphene, topology is not natural in non-Hermitian TBG systems. 

To answer these two questions, we add non-Hermitian perturbation on one of the twisted layers. 
As shown in Fig.~\ref{fig:fig1}(a), balanced gain and loss are put on the bottom layer (in red and blue, respectively).
The superposing of two graphene layers with a twist angle between them creates moir\'e patterns. In the mBZ shown in Fig.~\ref{fig:fig1}(b), the low-energy band structure consists of Dirac cones from the rotated layer located at the $K_M$ and $K'_M$ corners. When the non-Hermitian perturbation is turned on the bottom layer, Dirac points at the $K_M$ ($K'_M$) corner in the mBZ at the $K'$ ($K$) valley morph into exceptional rings and the corresponding Dirac cones become exceptional cones, as shown in Figs.~\ref{fig:fig1}(c) and \ref{fig:fig1}(d) respectively.  We find that this non-Hermitian system hosts special angles for which multiple exceptional cones coincide at $\Gamma_M$ points in an mBZ. We call this phenomenon ``exceptional magic". And importantly, our results demonstrated that the total Berry charge is conserved in the mBZ even after the merging of exceptional rings; this is a large difference from non-Hermitian single-layer graphene.

The paper is organized as follows. Section \ref{sec:model} introduces the non-Hermitian TBG model. Section \ref{sec:spectrum} shows the presence of exceptional rings in the real and imaginary parts of the low-energy moir\'e band spectrum. In Sec.~\ref{sec:tripod}, we use a non-Hermitian three-tripod model to show that the non-Hermitian chiral symmetry protects the exceptional rings. In Sec.~\ref{sec:NHmagic}, we will show the existence of a non-Hermitian flat band at an exceptional magic angle larger than the Hermitian magic angle.
In Sec.~\ref{sec:Chern}, we calculate the Berry curvature and the corresponding Chern number. Finally, we conclude our results in Sec.~\ref{sec:conclusion}.

\section{\label{sec:model}The non-Hermitian bilayer model}
We consider balanced gain and loss on the bottom layer of the bilayer system. We can write down a non-Hermitian generalization of the single-valley model for twisted bilayer graphene. The low-energy Hamiltonian for the mBZ at the $K$ or $K'$ valley is
\begin{equation}
    H=\begin{pmatrix}h_b(\bf{k}) & \bf{T}(\bf{r})\\\bf{T}^\dagger(\bf{r}) & h_t(\bf{k})\end{pmatrix},
\label{eq:H}
\end{equation}

where the top layer Hamiltonian is
\begin{equation}
h_{t}=-i\hbar v_F\bm{\sigma}_{\theta/2}\bm{\nabla}   
\end{equation} 
and the non-Hermitian bottom layer Hamiltonian is 
\begin{equation}
	h_{b}=-i\hbar v_F\bm{\sigma}_{-\theta/2}\bm{\nabla}+i\lambda_{V}\sigma_z.    
\end{equation}

 Here, $i\lambda_{V}\sigma_z$ is a balanced gain and loss imposing on the bottom layer, which induces non-Hermiticity, and $\bm{\sigma}_{\theta/2}$ corresponds to the rotated Pauli matrices $ e^{-i\theta\sigma_z/4}(\pm\sigma_x,\sigma_y) e^{i\theta\sigma_z/4}$, with $\pm$ for $K$ and $K'$ valley, respectively. $\textbf{T}(\textbf{r})=\sum^3_{j=1} T_j e^{-ik_\theta\textbf{q}_j\cdot\textbf{r}}$ is the sublattice-dependent moir\'e potential that couples the two layers. The interlayer coupling is in the form
\begin{equation}
T_{j+1}=w_0\sigma_0+w_1 (\sigma_x \cos j\phi+\sigma_y\sin j\phi)
\end{equation}
with $\phi=2\pi/3$ and $w_0(w_1)$ being the interlayer coupling strength in the AA (AB) region.
The unit vectors are $\textbf{q}_1=(0,-1)$ and $\textbf{q}_{2,3}=(\pm\sqrt{3}/2,1/2)$. $k_\theta=2k_D\sin(\theta/2)$ is the moir\'e modulation vector and $k_D=4\pi/(3a_0)$ is the magnitude of
the Dirac wave vector, where $a_0$ is the lattice constant of monolayer graphene. The hopping strengths $w_0$ and $w_1$ encode the inter-layer couplings in the AA and AB regions respectively. Correspondingly, dimensionless  parameters can be defined as $\alpha_i=\frac{w_i}{\hbar v_Fk_\theta}$ for $i=0,1$.

The generators of the magnetic space group $P6'2'2$ for the Hermitian twisted bilayer graphene include a $C_{3z}=e^{i\frac{2\pi}{3}\sigma_z}$ rotation symmetry, a $C_{2x}=\tau_x\sigma_x$ rotation symmetry, and a $C_{2z}T=\sigma_x\mathcal{K}$ symmetry~\cite{song2021twisted}. The single-valley Hamiltonian only has the $C_{3z}$ symmetry, since $C_{2z}$ and $T$ symmetries play the role of mapping one valley to the other. We can check that the non-Hermitian model preserves the $C_{3z}$ rotation symmetry.

If there is no AA stacking ($w_0=0$), the Hamiltonian preserves chiral symmetry $\mathcal{C}\mathcal{H}(\textbf{r})\mathcal{C}^\dagger=-\mathcal{H}(\textbf{r})$ under the chiral symmetry operator $\mathcal{C}=\sigma_z$ in the absence of non-Hermiticity~\cite{carr2019exact,tarnopolsky2019origin,song2021twisted}. 
In the presence of the balanced gain and loss $i\sigma_z$ term, the chiral symmetry is broken but the non-Hermitian chiral symmetry is preserved, which is crucial for the presence of non-Hermitian topology.
 
In the following numerical calculations, we take $w_1$ = 110 meV,  $\hbar v_F/a_0 = 2.413 eV$, and $w_0=0$ in the chiral limit.  In the absence of non-Hermiticity, the first magic angle is given by $\alpha_{1H} \approx 0.593$, which corresponds to $\theta\approx1.05$\degree.

\section{\label{sec:spectrum}Exceptional rings}

\begin{figure}[t]
\includegraphics[width=0.99\linewidth]{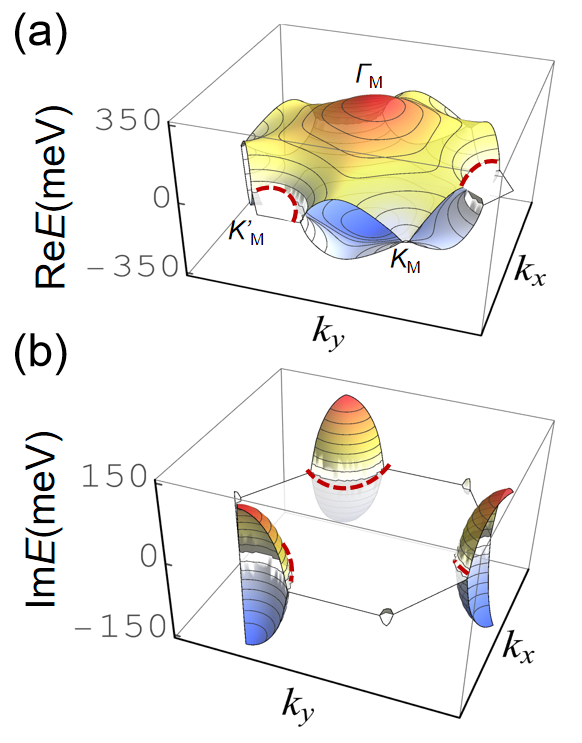}
\caption{(Color online)  In the presence of $\lambda_V$, exceptional rings appear. (a) The real part and (b) the imaginary part of the lowest-energy moir\'e band spectrum. The exceptional rings are marked by red line segments. The twist angle is taken as $\alpha_1=0.2$ and the non-Hermitian strength is $\lambda_V=0.07$ eV. 
 }
 \label{fig:fig2}
\end{figure}

Let us investigate the band structure for a twist angle away from the Hermitian magic angle. 
In the absence of non-Hermiticity, this is not a magic angle with a flat band. 
If the interlayer coupling effect is absent ($w_0=w_1=0$), the dispersion in the mBZ is equivalent to the folded dispersion in the Brillouin zone of graphene. Since the physics in the $K$ valley and in the $K'$ valley are related, we only discuss the $K$ valley in the following. The Dirac point at the $K$ corner in the top layer is mapped to $K_M$ point in the mBZ, while that at the $K$ corner of the bottom layer is mapped to $K'_M$ point in the mBZ.

In the presence of the non-Hermitian perturbation $\lambda_V$, the energy dispersion becomes complex. In Fig.~\ref{fig:fig2}(a), we can see that the non-Hermitian perturbation deforms the Dirac cones at $K'_M$ corners.  More specifically, the degeneracy point at the $K'_M$ points in the real part of the spectrum morphs into a ring of degeneracy points. This is a stark difference from single-layer graphene with on-site gain and loss. The latter has deformed Dirac cones in both $K$ and $K'$ corners in the Brillouin zone. This difference is due to the fact that the Dirac cones at the $K_M$ and $K_M'$ points in non-Hermitian TBG are folded from different graphene layers. The non-Hermitian $\lambda_V$ can only deform the Dirac cone on the bottom layer.

In particular, the degeneracy of the two lowest-energy states leads to the appearance of singularity points in momentum space. At the same time, a ring of degeneracy points appears at the same momentum position in the imaginary part of the spectrum, as shown in Fig.~\ref{fig:fig2}(b). We found that not only do the eigenvalues coalesce but also their corresponding eigenvectors coalesce at this point. This confirms that these singularity points are exceptional points in momentum space. Thus, the Dirac cone is turned into a ring of exceptional points.

Note that the spawning of exceptional ring out of a Dirac cone in non-Hermitian TBG is similar to that in other two-dimensional non-Hermitian semimetals protected by non-Hermitian chiral symmetry~\cite{zhen2015spawning,yoshida2019symmetry,okugawa2019topological,kawabata2019classification}. It is worthwhile to investigate the symmetry in non-Hermitian TBG.

 \section{\label{sec:tripod} Tripod model and non-Hermitian chiral symmetry}

To understand the emergence of non-Hermitian exceptional ring, we use the simplified tripod model~\cite{Bistritzer_2011}, which truncates Eq.~\eqref{eq:H} at the first honeycomb shell. The Hamiltonian is written as
\begin{equation}
    H_{tri}=\begin{pmatrix}
        h_b(\bf{k}) & T_{1} & T_{2} & T_{3} \\
         T_{1} & h_{t1}(\bf{k}-\bf{q}_1) & 0 & 0 \\
         T_{2} & 0 & h_{t2}(\bf{k}-\bf{q}_2) & 0 \\
          T_{3} & 0 & 0  & h_{t3}(\bf{k}-\bf{q}_3)         
    \end{pmatrix}
\end{equation}
where $h_{tj}$ for $j=1,2$, and $3$ is on the top layer and $h_b$ is on the bottom layer. Since $h_b$ is non-Hermitian, the Hamiltonian $H_{tri}$ has biorthogonal eigenvectors. In this system, the left and right eigenvectors are four two-component spinors: 
\begin{equation}
	\Psi_\alpha^T(\textbf{k})=(\psi_0(\textbf{k}),\psi_1(\textbf{k}),\psi_2(\textbf{k}),\psi_3(\textbf{k}))_\alpha^T, \quad \alpha=L,R.
\end{equation}

Using perturbation theory one can derive the effective Hamiltonian in the space of $\psi_0$.
The Hamiltonian $H_{tri}$ can be divide into a momentum-independent part $\mathcal{H}^{(0)}$ and a momentum-dependent part 
$\mathcal{H}^{(1)}_k$. The $\psi_{jR}$ can be expressed by
\begin{equation}
	\psi_{jR}=-h_j^{-1}T_j\psi_{0R}.
\end{equation}

Since $h_j$ is independent of non-Hermiticity, we can expect $T_jh_j^{-1}T_j=0$, which is similar to the Hermitian model. Thus, the $\psi_0$ spinor in the zero-energy eigenstates satisfies
\begin{equation}
	h_0\psi_{0R}=0.
\end{equation}
This form seems to be the same as that of the Hermitian TBG. However, $h_0$ is in a non-Hermitian form; the $\psi_{0R}$ is different from $\psi_{0L}$. We can check that the biorthorgonal wave functions are normalized as
\begin{equation}
	\langle\Psi_{0L}|\Psi_{0R}\rangle=1+3(\alpha_0^2+\alpha_1^2).
\end{equation}

The effective Hamiltonian matrix elements to leading order in $k$ is $\langle \Psi^{(i)}|\mathcal{H}^{(1)}_\textbf{k}|\Psi^{(j)}\rangle=\psi_0^{(i)\dag}(-v_F^*\rm{\sigma}\cdot\textbf{k}+i\lambda_V^*\sigma_z)\psi^{(j)}_0$, which gives the effective Hamiltonian
\begin{equation}
	\label{eq:effH}
H_\text{eff}=-v_F^*\rm{\sigma}\cdot\mathbf{k}+i\lambda_V^*\sigma_z, 
\end{equation}
with the renormalized velocity being
\begin{equation}
	\label{eq:revf}
	\frac{v_F^*}{v_F}=\frac{1-3\alpha_1^2}{1+3(\alpha_0^2+\alpha_1^2)}
\end{equation}
and the renormalized non-Hermitian strength being
\begin{equation}
	\frac{\lambda_V^*}{\lambda_V}=\frac{1-3(\alpha_0^2-\alpha_1^2)}{1+3(\alpha_0^2+\alpha_1^2)}.
\end{equation}
We can see that aside from a renormalized velocity and non-Hermitian strength, the form of Hamiltonian is identical to that of the continuum model Hamiltonian of non-Hermitian single-layer graphene.
The form of the renormalized velocity is the same as that of the Hermitian TBG. It decreases with increasing interlayer coupling in the AB region $\alpha_1$.

 We notice that the effective model in Eq.~\eqref{eq:effH} preserves the non-Hermitian chiral symmetry as
\begin{equation}
	\mathcal{C}H_\text{eff}^\dagger(\textbf{k})\mathcal{C}^{-1}=-H_\text{eff}(\textbf{k}),
\end{equation}
where the unitary operator $\mathcal{C}=\sigma_z$ satisfies $\mathcal{C}^2=1$. This non-Hermitian chiral symmetry is distinct from the Hermitian chiral symmetry due to $H(\textbf{k})\neq H^\dagger(\textbf{k})$~\cite{kawabata2019classification}.

The non-Hermitian chiral symmetry indicates that the eigenvalues come in pairs of ($E,-E^*$), which means the two eigenvalues $E_1$ and $E_2$ of $H_\text{eff}(\textbf{k})$ satisfy
\begin{equation}
	E_1=-E^*_2 \text{ or } E_i=-E^*_i\quad for\quad i=1,2.
\end{equation}
This ensures $\Delta=(E_1-E_2)^2$ is real, which is the discriminant of characteristic polynomial $f(E,\textbf{k})=det[E-H_\text{eff}(\textbf{k})]$. As we know, if both the real and imaginary parts of $\Delta$ are $0$, there is an exceptional point.
Since the constraint $Im(\Delta)=0$ is always satisfied in our system, the degrees of freedom (e.g., $k_x$ and $k_y$) outnumbers the only constraint $Re(\Delta)=0$, which leads to the presence of a higher-dimensional exceptional geometry --- the exceptional ring. Thus, the non-Hermitian chiral symmetry protects the exceptional ring in non-Hermitian TBG. A similar mechanism can be found in other two-dimensional non-Hermitian semimetals protected by non-Hermitian chiral symmetry~\cite{zhen2015spawning,yoshida2019symmetry,okugawa2019topological,kawabata2019classification}.\\

\begin{figure}[t]
	\includegraphics[width=0.95\linewidth]{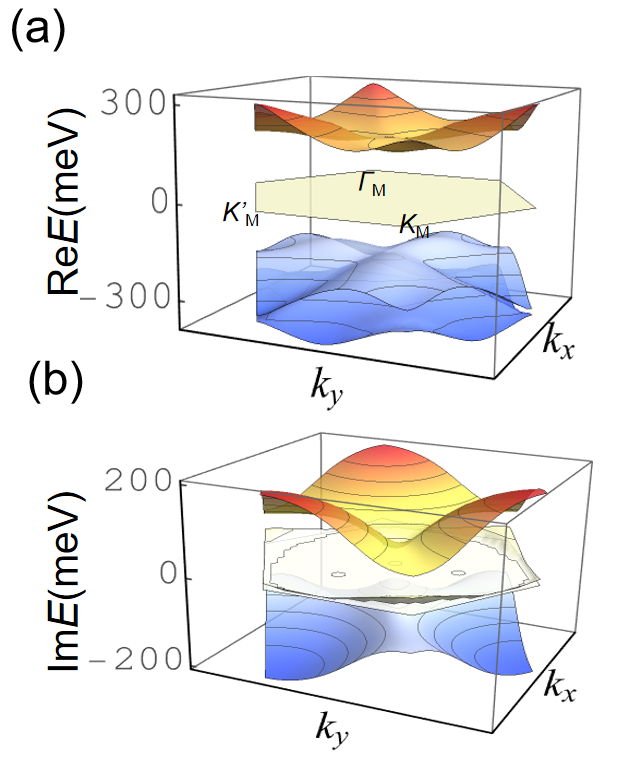}
	\caption{(color online)   At an ``exceptional magic angle" $\alpha_{1c}$, the lowest band shows (a) a real part that is extremely flat and (b) an imaginary part with a finite lifetime. The non-Hermitian strength is taken as $\lambda_V=0.07$eV. The exceptional magic angle is $\alpha_{1c}=0.46$ away from the Hermitian magic angle $\alpha_{1H}\approx0.593$.
	} 
	\label{fig:fig3}
\end{figure}	

\begin{figure}[t]
	\includegraphics[width=0.95\linewidth]{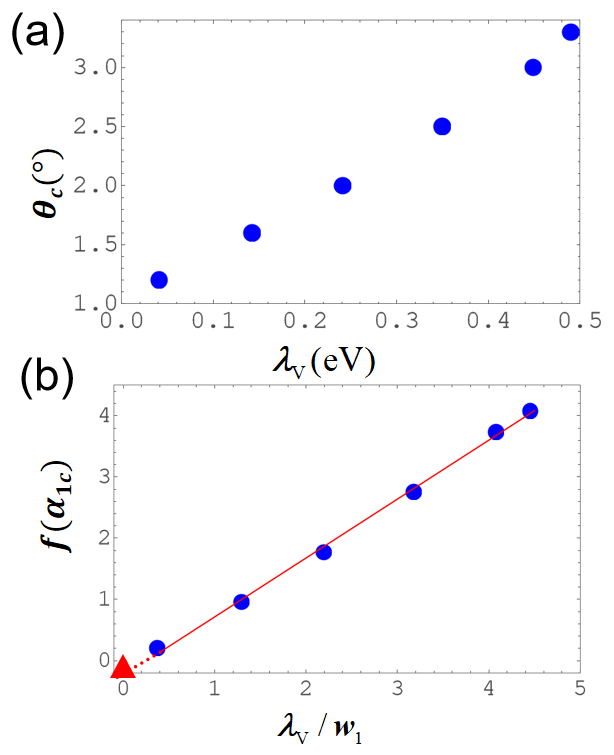}
	\caption{(color online)  (a) The exceptional magic angle $\theta_c$ as a mononic increasing function of non-Hermitian strength $\lambda_V$ (b) $f(\alpha_{1c})$ as a function of the ratio between the non-Hermitian strength $\lambda_V$ and the interlayer hopping $w_1$. The red line is the linear fit with a slope of 0.96. The red triangle marks $f(\alpha_{1H})$ in the Hermitian limit.
	} 
	\label{fig:fig4}
\end{figure}

	\section{\label{sec:NHmagic} Non-Hermitian flat band at exceptional magic angle}
	In the previous sections, we have shown that an exceptional ring appears at a non-magic twist angle. Here, we show the emergence of a non-Hermitian flat band at an ``exceptional magic angle" and discuss the relationship between the exceptional magic angle and non-Hermitian strength considering Fermi velocity renormalization.
	
	For a fixed non-Hermitian strength $\lambda_V$, we expect that the exceptional rings cover more regions in the mBZ with decreasing twist angle. This is because the size of the mBZ, $k_\theta=2k_D\sin(\theta/2)$, decreases with decreasing $\theta$.  As the twist angle decreases to a critical value, the three exceptional rings centered at different $K'_M$s cross each other at $\Gamma_M$ and cover the whole mBZ. As shown in Fig.~\ref{fig:fig3}, we find that the whole real part of the two lowest bands becomes $0$. This means the flat bands are compatible with non-Hermiticity. The presence of these non-Hermitian flat bands in moir\'e system is the main finding in this work. We can see that these non-Hermitian flat bands are isolated from higher-energy bands, which are distinguished from non-Hermitian flat bands in geometry-frustrated lattices.
	
	 The corresponding critical twist angle can be denoted as the ``exceptional magic angle", analogous to the magic angle in the Hermitian case. This exceptional magic angle can be defined by the twist angle at which the bandwidth of the lowest band is minimized. While the bottom of the lowest band at $K'_M$ from the Dirac point is always at zero energy, the band top should be minimized at the exceptional magic angle. Thus, the band top at the $\Gamma_M$ point takes the zero energy value at the exceptional magic angle $\theta_c$ and dimensionless $\alpha_{1c}$

     \begin{equation}
     	E=\sqrt{(\hbar v^*_Fk_{\theta c})^2-\lambda_V^{*2}}=0.
     	\label{eq:critical}
     \end{equation}
			
	In Fig.~\ref{fig:fig4}(a), we can see that the exceptional magic angle $\theta_{c}$ increases with increasing non-Hermitian strength. This is consistent with our expectation that exceptional rings cover more regions in the mBZ with decreasing the twist angle. Importantly, this offers a feasible way to realize a non-Hermitian flat band at a large twist angle, which is easier to be precisely controlled than at a Hermitian magic angle.

	Taking into account the renormalization effect of the Fermi velocity, the relationship between $\alpha_{1c}$ and $\lambda^*_V$ can be established by using Eqs.~\eqref{eq:revf} and \eqref{eq:critical}. We find that 
	\begin{equation}
		\label{eq:fit}
		\lambda^*_V=w_1 f(\alpha_{1c}), \quad\text{with } f(\alpha_{1c})=\frac{1-3\alpha_{1c}^2}{(1+3\alpha_0^2+3\alpha_{1c}^2)\alpha_{1c}}.
	\end{equation}
	In the chiral limit $(\alpha_0=0$), $\lambda^*_V$ can be replaced by $\lambda_V$. In Fig.~\ref{fig:fig4}(b), we plot $f(\alpha_{1c})$ as a function of $\lambda_V/w_1$. A linear fit with a slope close to 1 confirms Eq.~\eqref{eq:fit}. Also, when $\lambda$ is close to $0$, $f(\alpha_{1c})$ approaches  $f(\alpha_{1H})$ with a Hermitian magic angle. This is consistent with the Hermitian limit.
	
	Note that when the size of mBZ is large, higher-order momentum terms will give obvious deviation from Eq.~\eqref{eq:fit}, since only the leading order in $k$ is considered in Eq.~\eqref{eq:revf}. In addition, the band crossing with a higher-energy dispersive band presented at large non-Hermiticity will break the isolation of the non-Hermitian flat band. Both conditions constrain the valid range of non-Hermitian strength.

\section{\label{sec:Chern} Berry Curvature and Chern number}
So far, we have shown the existence of exceptional geometries in the non-Hermitian TBG. In this section, we explore their topological properties numerically.
\begin{figure}
	\vspace*{0.5cm}
	\includegraphics[width=0.8\linewidth]{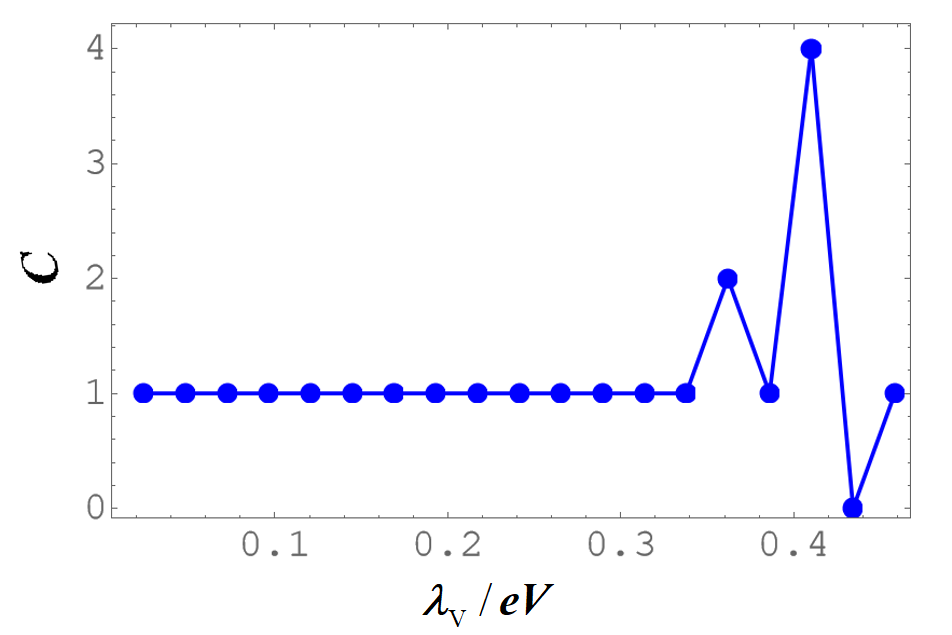}
	\caption{(color online) Topological charge for the $\sigma_z=1$ band in the mBZ as a function of non-Hermitian strength.  A small in-plane electric field is imposed to open the gap. The results are in the chiral limit $\alpha_0=0$ and the twist angle is taken as $\alpha_1=0.384$. While the topological charge remains to be quantized when the non-Hermitian strength exceeds the critical value $\lambda_{Vc}=0.15$ at which the non-Hermitian flat band appears, it fluctuates when the $\sigma_z=1$ band crosses with higher bands at large non-Hermitian strength above $\lambda_V=0.3$ eV.
		\label{fig:berry}}
\end{figure}

 The topological stability is ensured by the nontrivial topology for a line gap, when a line can be drawn on the complex energy plane to separate the energies of two bands along a closed path. In our system, the effective Hamiltonian, Eq.\eqref{eq:effH}, preserves the $PT$ symmetry  $(PT)H_\text{eff}(\textbf{k})(PT)^{-1}=H_\text{eff}(\textbf{k})$ with $PT=\sigma_x K$. In addition, symmetry-protected exceptional rings with the exceptional geometry dimension $d_\text{EP}=1$ in $d=2$ spatial dimension have codimension $p$ defined as $p=d-d_{EP}=1$.
From the non-Hermitian topology theory~\cite{kawabata2019classification}, there is a well-defined Chern number corresponding to the $\mathbb{Z}$ index for a line gap for the real part of the energy.
	
The topological charge or Chern number of an exceptional contour can be obtained by integrating the Berry connection $A(\mathbf{k})$ along a closed surface containing the exceptional contour
 
\begin{equation}
    \mathcal{C}=\frac{1}{2\pi}\oint_{\partial S} \mathbf{A}(\mathbf{k})\cdot d\mathbf{k}= \frac{1}{2\pi}\int_{S}\Omega(\mathbf{k})\cdot d\mathbf{S},
    \label{eq:Chern}
\end{equation}
where the Berry connection is defined by the eigenvectors as
\begin{equation}
\mathbf{A}^{\beta,\beta'}=i\langle\psi^{\beta}(\mathbf{k})|\nabla_\mathbf{k}|\psi^{\beta'}(\mathbf{k})\rangle.   
\end{equation}
Here, $|\psi^{\beta'}(\mathbf{k})\rangle$ and $\langle\psi^{\beta}(\mathbf{k})|$ can be the left or right eigenvectors. Since previous study has proven that the total topological charge is the same for four possible choices of eigenvectors~\cite{shen2018topological}, we choose $\beta=L$ and $\beta'=R$ in the following calculation.

In numerical calculation, the Berry connection is presented by the local Berry curvature $\Omega(\mathbf{k})=\nabla_\mathbf{k}\times\mathbf{A}(\mathbf{k})$, which can be calculated as
\begin{widetext}
\begin{equation}
\Omega(\mathbf{k})^{\beta,\beta'}=\lim_{q\rightarrow0}\frac{1}{4q^2}	\langle\psi^{\beta'}_{\mathbf{k}-q\hat{x}-q\hat{y}}|\psi^{\beta}_{\mathbf{k}-q\hat{x}+q\hat{y}}\rangle
\langle\psi^{\beta'}_{\mathbf{k}-q\hat{x}+q\hat{y}}|\psi^{\beta}_{\mathbf{k}+q\hat{x}+q\hat{y}}\rangle
\langle\psi^{\beta}_{\mathbf{k}+q\hat{x}+q\hat{y}}|\psi^{\beta'}_{\mathbf{k}+q\hat{x}-q\hat{y}}\rangle
\langle\psi^{\beta}_{\mathbf{k}+q\hat{x}-q\hat{y}}|\psi^{\beta'}_{\mathbf{k}-q\hat{x}-q\hat{y}}\rangle,
\label{eq:Berry}
\end{equation} 
\end{widetext}
with $q$ being half of the lattice constant.

Upon summing the Berry curvature on the mBZ, we can obtain the total Berry charge or Chern number
\begin{equation}
	\mathcal{C}=\sum_{mBZ}\Omega^{LR}(\textbf{k}).
\end{equation}

 The non-Hermitian perturbation transforms a Dirac point into an exceptional ring. In the mBZ, there are three-thirds of exceptional rings from Dirac points in different moir\'e corners. For small $\lambda_V$, we can find that the total Berry charge in one mBZ is 1. This can be easy to understand from two facts. One is that each of the three Dirac points corresponds to Berry curvature monopoles of charge 1 in the absence of non-Hermiticity. The other is that the topological charge is preserved on the exceptional contour that forms from the original Dirac point. Thus, we can expect that each of the three exceptional contours contributes one-third of the Berry charge to one mBZ.
 
Most interestingly, as shown in Fig.~\ref{fig:berry}, even when $\lambda_V$ exceeds the critical non-Hermitian strength $\lambda_{Vc}=w_1f(\alpha_c)$, which corresponds to the exceptional magic angle $\alpha_c$, the total Berry charge remains $1$. This is opposite to the merging of exceptional contours with opposite topological charge~\cite{cerjan2018effects}, in which the topological charge is dissipated and gives a single, uncharged exceptional contour. This indicates a breakdown of the Nielsen-Ninomiya theorem and distinguishes the valley topological features of non-Hermitian TBG from a single-layer graphene.  This show that the non-Hermitian flat band is topological nontrivial, which distinguishes it from the non-Hermitian flat bands in geometry-frustrated lattices~\cite{ge2015parity,chern2015pt,leykam2017flat,qi2018defect,zhang2019flat}

Note that the band crossings with the remote bands will lead to a fluctuation of topological number as the bands are degenerated. This gives an upper limit of $\lambda_V$ for topological non-Hermitian flat bands.

\section{\label{sec:conclusion} Discussion and conclusion}
In this paper, we have studied a non-Hermitian generalization of twisted bilayer graphene system with balanced gain and loss on one of the layers. 
We found that exceptional rings are centered at the $K_M$ ($K'_M$) corner in the moir\'e Brillouin zone in the $K'$ ($K$) valley. This is due to the folding of the Brillouin zone, and it is absent in the single-layer graphene.  We found that the exceptional rings are protected by non-Hermitian chiral symmetry.
The exceptional rings coincide and form bands whose real parts are flat at a specific non-Hermitian strength, which is denoted as ``exceptional Dirac magic". This non-Hermitian flat band is characterized with a robust Chern number, $\mathcal{C}=1$, in moir\'e Brillouin zone.

The presence of non-Hermitian flat bands at an ``exceptional Dirac magic" angle introduces a new type of band engineering for twisted moir\'e systems. It is known that the electronic properties of bilayer graphene near a magic twist angle are extraordinarily sensitive to the carrier density and to controllable environmental factors such as 
the proximity of nearby gates and twist-angle variation~\cite{andrei2020graphene}. The presence of non-Hermiticity increases the magic twist-angle value and 
provides some degrees of tuning capability that can reduce the severity of absolute twist-angle control requirements.

The non-Hermitian flat bands in moir\'e systems are distinguished from those in geometrical frustration lattices, which are embedded in dispersive bands~\cite{leykam2017flat}. While the non-Hermitian flat bands in frustrated lattices have been shown to give compact localized states, the wave propagation of non-Hermitian flat bands in moir\'e system can also be investigated in cold atom and metamaterial systems.
Recently, an ultracold atom experiment has had a breakthrough in the realization of atomic Bose-Einstein condensate in twisted bilayer optical lattices ~\cite{meng2023atomic}.  Also, following the proposals on the photonic analog~\cite{oudich2021photonic,lou2021theory,sunku2018photonic,deng2020magic,valagiannopoulos2022electromagnetic} and the phononic analog~\cite{oudich2022twisted,gardezi2021simulating} of twisted bilayer graphene, demonstration of optical bilayer photonic crystal devices has been recently reported in the microwave range~\cite{lou2022tunable} and the optical frequency range~\cite{tang2023experimental}.
Since on-site dissipation is experimentally realizable in cold-atom and metamaterial systems ~\cite{ding2022non,ashida2020non,bergholtz2021exceptional}, the realization of non-Hermitian flat bands in these systems can be anticipated. In particular, the Weyl exceptional ring has been realized in photonic experiments~\cite{cerjan2019experimental,song2023observation}.

The non-Hermitian flat band is topological and may lead to the realization of non-Hermitian fractional quantum Hall states. First, non-Hermiticity can enhance strong correlated phenomena. For example, a numerical study on non-Hermitian interacting system on the honeycomb lattice shows that non-Hermitian enhances the antiferromagnetic ordered phase and changes its transition to Dirac semimetal~\cite{yu2024non}. 
Second, non-Hermiticity may induce a topological phase that does not have its Hermitian counterpart.
For example, the study of non-Hermitian quantum Hall states finds that the best quantization of the non-Hermitian topological invariant is observed at high carrier density and low field~\cite{ozer2024non}, which provide efficient operation of devices.

Note added. Upon completion of our manuscript, we became aware of a recent paper on non-Hermitian twisted bilayer graphene, in which non-Hermiticity is on the hopping amplitude and topological aspects are undiscussed~\cite{esparza2024exceptional}.

\begin{acknowledgments}
We thank the helpful discussion with Zhen-Yu Zhang. This work is supported by the National Natural Science Foundation of China (Grants No.~12104099 and No.~12274095) and the Guangzhou Science and Technology Program (Grant No.~2024A04J0272).
\end{acknowledgments}

\bibliography{ref}

\end{document}